\documentclass[reprint,aps]{revtex4-1}

\usepackage{graphicx}
\usepackage{dcolumn}
\usepackage{bm}
\usepackage{epstopdf}
\usepackage{hyperref}
\usepackage[export]{adjustbox}
\usepackage{lipsum}
\usepackage{amsmath, amsfonts, amssymb}
\begin{document}


\title{Wettability Modulated Charge Inversion and Ionic Transport in Nanofluidic Channels}
\author{Vaseem Akram Shaik$^{1}$, Aditya Bandopadhyay$^{2}$, Syed Sahil Hossain$^{1}$, and Suman Chakraborty$^{1,2}$}
	\email{suman@mech.iitkgp.ernet.in}
\affiliation{$^{1}$ Department of Mechanical Engineering, Indian Institute of Technology, Kharagpur-721302, India}
\affiliation{$^{2}$ Advanced Technology Development Center, Indian Institute of Technology, Kharagpur-721302, India}
\date{\today}

\begin{abstract}

We unveil the role of substrate wettability on the reversal in the sign of the interfacial charge distribution in a nanochannel in presence of multivalent ions. In sharp contrast to the prevailing notion that hydrophobic interactions may trivially augment the effective surface charge, we demonstrate that the interplay between surface hydrophobicity and interfacial electrostatics may result in a decrease in the effective interfacial potential, and a consequent charge inversion over regimes of low surface charges. We also show that this phenomenon, in tandem with the interfacial hydrodynamics may non-trivially lead to either augmentation or attenuation or even reversal of the net streaming current, depending on the relevant physical scales involved. These results, supported by Molecular Dynamics simulations and experimental data, may bear far ranging consequences in understanding complex biophysical processes and designing nanofluidic devices and systems involving multivalent counterions.

\begin{description}
\item[PACS numbers]
\end{description}
\end{abstract}

\maketitle


A charged substrate in contact with a multivalent ionic solution may attract larger numbers of ions possessing opposite sign (counterions) than that necessary to screen its surface charge. 
This counterintuitive phenomenon, leading to a flip in the sign of the interfacial charge, is commonly known as charge inversion (CI) \cite{Grosberg2002}. CI plays a vital role in biological processes such as DNA condensation \cite{Besteman2007}, viral packaging \cite{Nguyen2013}, and drug delivery \cite{Li2008,*Gerelli2008}. 
CI has also been found to be responsible for the reversal in the sign of the ionic current due to pressure-driven advection (streaming currents) in the presence of an electrically charged interfacial layer adhering to the substrate \cite{vanderHeyden2006}. 
Besides this, CI is also responsible for an inverted mobility of charged species at high bulk ionic concentrations \cite{Jimnez2012,*Semenov2013}. 

Theoretically, the apparent anomalous behavior associated with CI has been related to nontrivial ion-ion correlations, as verified through molecular scale simulations \cite{Boda2002,Qiao2004,Labbez2009,Gillespie2011}. 
However, the reported theories on CI \cite{Outhwaite1980,*Outhwaite1982,*Outhwaite1983,BariBhuiyan2004,Bazant2011,Storey2012} have trivially considered the substrate to be perfectly wetting in nature, which is unlike the characteristic of real biological and engineering systems. As a consequence, the role of substrate wettability on charge inversion remains poorly understood. Such inadequacies in theoretical perception possibly
stem from the complexities in simultaneously capturing the coupled electrostatics of CI and the underlying wettability-induced interfacial phenomena over the relevant spatio-temporal scales.

In this work, we bring out the inconspicuous role of wettability-modulated CI on electrokinetic phenomena in nanofluidic channels. 
In contrast to the earlier reported findings on monovalent ions that hydrophobic interactions lead to an enhanced effective zeta potential \cite{Joly2006,Joly2004}, we show that for multivalent ions, reduction in the substrate wettability may lead to a decreased effective zeta potential, and even CI at a relatively low surface charge density. 
In addition, we demonstrate that the interplay between hydrodynamics and interfacial electromechanics for multivalent counterions may non-trivially lead to either reduction (including inversion) or augmentation of the streaming current, depending on the physical scales involved. Our theoretical model devised on phase field formalism is shown to be in excellent agreement with molecular dynamics (MD) simulations as well with reported experimental data \cite{vanderHeyden2006}, in an effort to support these unexpected trends.

\begin{figure}[t]
\includegraphics[bb=35 646 312 760,clip=true,scale=0.8]{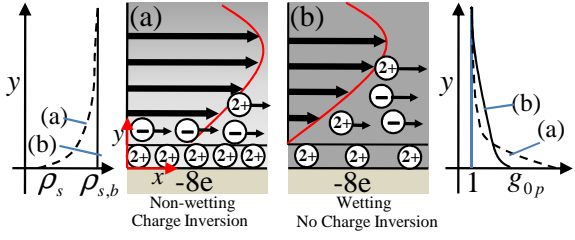}
\caption{\label{fig:fig_1} (color online) Schematic depicting the two scenarios: the wall is (a) non-wetting and (b) perfectly wetting. For case (a), the solvent depletion (represented by the gradient in the background color) leads to a local variation in the solvent density (leftmost plot; the dashed line depicts the non-wetting case where there is a depleted wall-adhering phase due to hydrophobic interactions). The consequent change in the ionic distribution, $\left( {{g_{0p}}} \right)$ is shown in the rightmost plot. Owing to lower solvent density, there is a larger counterion concentration closer to the wall for case (a), denoted by the dashed line. CI is more strongly manifested in case (a) as compared to case (b). Notably, over length scales of few nanometers, a larger ionic flux in presence of an external pressure-gradient driven flow is also obtained for case (a), owing to a less-dense phase which leads to slipping hydrodynamics over interfacial scales with excess mobile charges.}
\end{figure}

Towards establishing the interplay of hydrophobic interactions and CI, we first refer to the schematic depicted in Fig.~\ref{fig:fig_1},
which qualitatively represents the solvent density, ionic distribution and ionic transport characteristics in the near-wall region of a channel containing multivalent ions. In our modeling paradigm, we couple the electromechanics with the physics of hydrophobic interactions over interfacial scales by relating the potential distribution $(\psi)$ in the wall-adjacent layer with the solvent density distribution through the Poisson equation: $\frac{d}{{dy}}\left( {\varepsilon \left( y \right)\frac{{d\psi }}{{dy}}} \right) =  - \sum\limits_i {{z_i}en_i^0{g_{0i}}(y)} ,\,\,{\rm{for}}\,\,y \ge a/2$, where {\it{y}} is the wall-normal coordinate, {\it{a}} is the diameter of hydrated ions (assumed same for both the ions), {\it e} is protonic charge, ${z_i}$ and ${n_i^0}$ are the valency and bulk number density of ions of {\it i}$^{th}$ species, ${g_{0i}}\left( y \right)$ is a modified wall-ion distribution function for the {\it i}$^{th}$ species (see \cite{[{See Appendix}]SI} for a discussion on the various interactions affecting $g_{0i}$); $\varepsilon$ is the permittivity of the medium which varies spatially in accordance to the relative phase distribution originating out of hydrophobic interactions \cite{Gongadze2011,*Bandopadhyay2012,*Gongadze2013}. 
The near-wall solvent density variations due to hydrophobic interactions, steric effects and ion-ion correlations \cite{BariBhuiyan2004} are simultaneously accounted by a modification of the Boltzmann description of the number density of the ions , so that: ${g_{0i}}(y) = \frac{{\rho \left( y \right)}}{{{\rho _l}}}{\zeta _i}(y)\exp \left( { - \beta {z_i}eL(\psi ) - \frac{{\beta z_i^2{e^2}}}{{8\pi \varepsilon \left( y \right)a}}\left( {F - {F_0}} \right)} \right)$, where $L(\psi )$ and $F$ are functions of potential ($\psi$) and transverse coordinate respectively, ${\zeta _i}\left( y \right)$ is the exclusion volume term of the {\it i}$^{th}$ ion, whereas ${F_0} = 1/(1 + {\kappa _0}a)$, and for a binary electrolyte solution, ${\kappa _0} = \mathop {\lim }\limits_{y \to \infty } \kappa  = \sqrt {{\textstyle{{{n^0}{e^2}\left( {{z_p} - {z_n}} \right)} \over {{\varepsilon _l}{k_b}T}}}}$. 
Here $\kappa$ is the local Debye-Huckel parameter given by, ${\kappa ^2} = \left( {{n^0}{e^2}/\left( {\varepsilon \left( y \right){k_b}T} \right)} \right)\left( {{z_p}{g_{0p}} - {z_n}{g_{0n}}} \right)$, $\beta  = {\left( {{k_b}T} \right)^{ - 1}}$, ${k_b}$ being the Boltzmann constant, $T$ being the absolute temperature, and $n^0 = {z_p}n_p^0 = -{z_n}n_n^0$, where $n_p^0$, $n_n^0$ and $z_p$, $z_n$ are the bulk number densities and valency of cations and anions respectively \cite{Outhwaite1982,Outhwaite1983}, $\rho \left( y \right)$ is the near wall solvent density distribution, and $\rho_l$ is the bulk solvent density. 
The Poisson equation, as mentioned above, is solved in conjunction with the equation $\frac{d}{{dy}}\left( {\varepsilon \left( y \right)\frac{{d\psi }}{{dy}}} \right) = 0$ for $y<{a}/{2}$ with the following matching boundary conditions for obtaining $\psi (y)$: (i) continuity of $\psi$ and $\varepsilon {\textstyle{{d\psi } \over {dy}}}$ at $y = a/2$, (ii) known surface charge density at the wall, $\sigma  =  - {\left. {\left( {\varepsilon \left( y \right)d\psi /dy} \right)} \right|_{y = 0}}$ and (iii) symmetry boundary condition at the centerline of the channel, ${\left. {\left( {d\psi /dy} \right)} \right|_{y = h}} = 0$, where {\it h} is the half channel height.

In order to mathematically close the above set of equations, we relate the variations in $\varepsilon \left( y \right)$ and $\rho \left( y \right)/{\rho _l}$ with the distribution of an order parameter $\phi$, by appealing to the phase field formalism \cite{Andrienko2003}. Here $\phi  = ({n_v} - {n_l})/({n_v} + {n_l})$, where $n_i$ represents the number density of the {\it i}$^{th}$ phase \cite{Andrienko2003}. $\phi  =  - 1$ represents the bulk liquid phase (denoted by subscript {\it l}), and $\phi  = 1$ (denoted by subscript {\it v}) represents the wall-adhering low density phase formed due to hydrophobic effects \cite{Chakraborty2007}. 
The calculation of equilibrium $\phi$ starts by first considering a free-energy functional which represents the excess Ginzburg-Landau-like free energy for a binary mixture given by \cite{Cahn1977}: $\Delta \Omega \left( \phi  \right) = \int {\left[ {\frac{k}{2}{{\left( {\frac{{d\phi }}{{dy}}} \right)}^2} + \Delta \omega \left( \phi  \right)} \right]dy}  + {\Omega _S}$, where $\Delta\omega(\phi)$ is the bulk free energy having a double well potential profile \cite{Chakraborty2007,Badalassi2003}, $\Delta \omega  = {\textstyle{B \over 4}}{\left( {{\phi ^2} - 1} \right)^2}$ and $\Omega_S$ is the surface energy that takes into account the interactions between the substrate and the fluid \cite{Andrienko2003}. Here $B$ is a positive constant such that $B\sim{k_bT_C}$ with $T_C$ being the critical temperature for the liquid-vapor coexistence, whereas ${\textstyle{k \over 2}}{(d\phi /dy)^2}$ is the interfacial energy with $k$ being a positive constant \cite{Cahn1977}. 
The minimization of free energy functional results in the Euler-Lagrange equation: $\frac{{d\Delta \omega }}{{d\phi }} - \frac{d}{{dy}}\left[ {\frac{k}{2}\frac{d}{{d\phi '}}\left\{ {{{\left( {\frac{{d\phi }}{{dy}}} \right)}^2}} \right\}} \right] = 0$ \cite{Chakraborty2012}. The interfacial value of $\phi$ is related to the surface wettability through the contact angle, which is given by $\cos {\theta _w} = \frac{{\phi _S^3 - 3{\phi _S}}}{2}$ \cite{Chakraborty2007,Chakraborty2008}. Combining the double well potential for $\Delta\omega$ with the Euler-Lagrange equation yields the following governing differential equation for the order parameter variable in a dimensionless form: $\frac{{{d^2}\phi }}{{d{{\bar y}^2}}} - {C^2}({\phi ^2} - 1)\phi  = 0$, where $C = a/\xi$ ($k\sim B{\xi ^2}$; $\xi$ being the multiple of interfacial thickness \cite{Chakraborty2007}), and $\bar y = y/a$. The corresponding boundary conditions are: ${\left. {\left( {d\phi /d\bar y} \right)} \right|_{\bar y = h/a}} = 0$ and ${\left. \phi  \right|_{\bar y = 0}} = {\phi _S}$. Properties, such as $\rho$ and $\varepsilon$ are interpolated as: $\chi  = {\chi _v}\frac{{1 + \phi }}{2} + {\chi _l}\frac{{1 - \phi }}{2}{\rm{ }}$, where $\chi$ is a generic property.
The  value of parameters and properties used in the numerical simulation are ${z_p}:{z_n} = 2:-1$, ${\varepsilon _l} = 78.5{\varepsilon _0}$, where ${\varepsilon _0}$, is the permittivity of the vacuum, ${\varepsilon _v}/{\varepsilon _l} = 1/15$ \cite{Gongadze2011,Bandopadhyay2012,Gongadze2013}, ${\rho _v}/{\rho _l} = 2.306 \times {10^{ - 5}}$, $T = 298$ K, $\xi  = 0.38$ nm \cite{Chakraborty2007,Lum1999}. The values of the remaining parameters are stated later.

When the ionic solution, energetically described as above, is driven by an external pressure gradient applied orthogonal to the direction of the variation of $\phi$ and $\psi$, one may have an effective axial electrical body force even though no external axial electric field is applied. This body force is because of the development of a back electromotive potential (also known as streaming potential) in an otherwise pressure-driven flow field, and may be described as: ${{\bf{F}}_{{\rm{EK}}}} = {\rho _e}{{\bf{E}}_{{\rm{s}}}}$, where $\rho_e$ is the electrical charge density and ${{\bf{E}}_{{\rm{s}}}}$ is the streaming electric field. This undetermined streaming field is obtained by setting the net ionic current, which is the sum of the streaming (advection) current and electromigration current, to be zero in presence of an applied pressure gradient \cite{Hunter1981,Kirby2010}.

In Fig. 2(a),
\begin{figure}[htb]
  \begin{minipage}[b]{0.8\linewidth}
    \includegraphics[width=.99\textwidth]{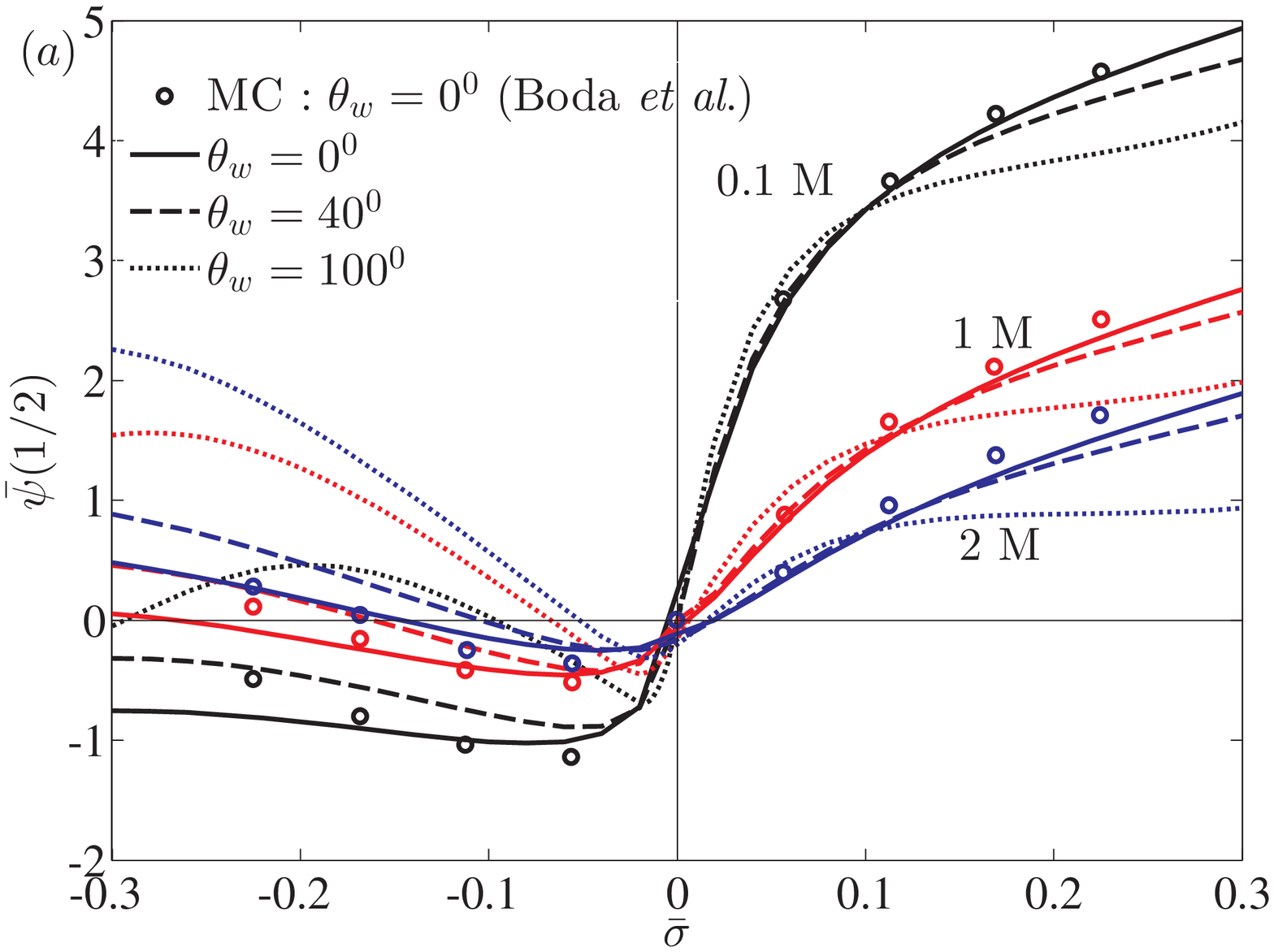}
    \label{fig:fig_2a}
  \end{minipage}\\
  \begin{minipage}[b]{.8\linewidth}
    \includegraphics[width=.99\textwidth]{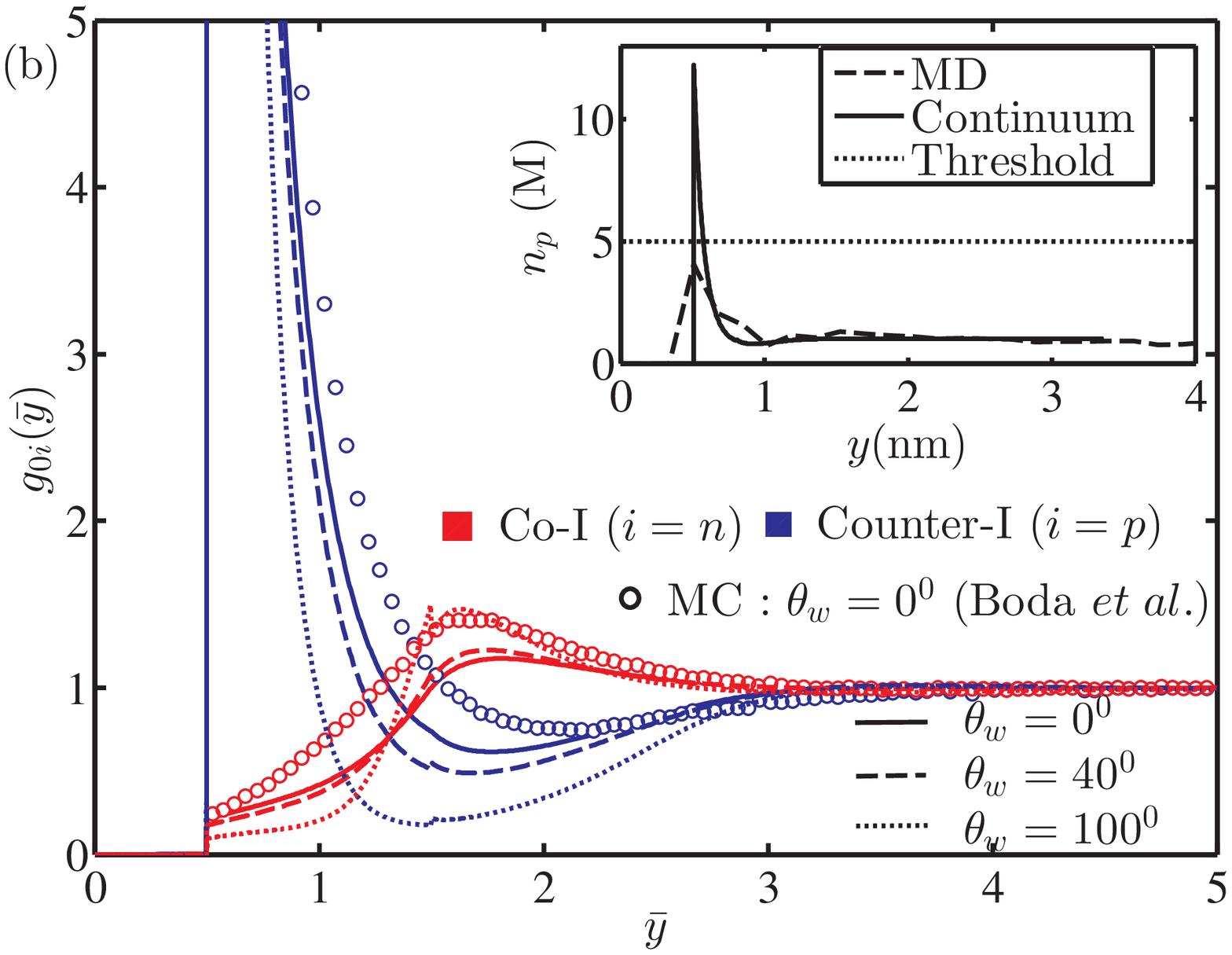}
    \label{fig:fig_2b}
  \end{minipage}  
\caption{\small(color online). (a) Reduced potential at location `{\it a}/2', $\bar{\psi}(1/2) = \beta e\,\psi \left( {a/2} \right)$ as a function of reduced surface charge density, $\bar{\sigma}  = \sigma {a^2}/e$ for a 2:1 binary electrolyte at various electrolyte concentrations and contact angles. (b) Singlet wall-ion density distribution functions for 2:1 binary electrolyte with a concentration of 1M and $\bar{\sigma}  =  - 0.1685$ at different contact angles. Comparison is made against MC results reported in \cite{Boda2002} (open circles). Inset depicts the counterion concentration spatial variation obtained from the present model and MD simulations (see \cite{SI} for details pertaining to MD simulations). It also depicts a concentration threshold (same as maximum counterion concentration in MD), above which the ions are considered to be jammed and immobile. Here, Co-I and Coun-I represent the coion and counterion wall-ion distribution functions respectively. The hydrated ionic diameter is considered as 0.3 nm according to the MC simulations \cite{Boda2002}.}
	\label{fig:2}
\end{figure}
we depict the variation in the effective surface potential ${\left. {\left( {\bar \psi  = \beta e\psi } \right)} \right|_{\bar y = 0.5}}$ vs dimensionless surface charge density, $\bar \sigma  = \sigma {a^2}/e$, for a $2:1$ electrolyte solution at various concentrations and contact angles (see \cite{SI} for a detailed description on the theoretical treatment of the modified Poisson-Boltzmann formalism). Additionally, in Fig. 2(b), we plot wall-ion distribution functions for $2:1$ electrolyte solutions at contact angles of 0$^0$, 40$^0$ and 100$^0$. We compare these results against reported Monte Carlo (MC) simulations for a perfectly wetting substrate \cite{Boda2002}. Considering reported experiments on nanochannel made of fused silica \cite{vanderHeyden2006}, we have chosen the contact angle, depending on the relative fraction of Si-OH and Si-O-Si groups on the surface dictated by the fabrication technique, to vary between 0$^0$ and 40$^0$ \cite{Lamb1982,*Israelachvili1989}. In addition to these contact angles, we also show the plots obtained for ${\theta _w} = {100^0}$, considering the emerging trends of using highly hydrophobic polymeric substrates for fabricating miniaturized fluidic channels \cite{Tandon2008}.

We refer our results in perspective of the potential at the plane located at `$a$/2'(shear plane), also known as the zeta potential \cite{Hunter1981}. For a perfectly wetting substrate $\left( {{\theta _w} = {0^0}} \right)$, the zeta potential calculated using the theory is in good agreement with MC simulation results \cite{Boda2002,BariBhuiyan2004}. When the counterions are monovalent, (positive half of Fig. 2(a)), two trends of $\bar{\psi}(1/2)$ vs $\bar{\sigma}$ are evident for an increasing contact angle. First, for low $\bar{\sigma}(0 < \bar \sigma  < 0.1)$, the zeta potential increases with the contact angle; a phenomenon previously noted as well  \cite{Joly2006,Joly2004}. Second, for high $\bar{\sigma}(0.1 < \bar \sigma  < 0.3)$, the zeta potential decreases with the contact angle as attributable to CI at the high surface charge densities. However, when the counterions are divalent (negative half of Fig. 2(a)), the magnitude of zeta potential decreases with the contact angle, eventually flipping its sign, after which the magnitude of this inverted zeta potential increases with the contact angle. All these observations indicate CI. More importantly, the substrate wettability, in conjunction with the ionic valency and substrate charge density, plays an important role in modulating the CI.

\begin{table*}[t]
\caption{\label{tab:table_1}\small Variation of streaming currents per unit width (from both MD and present model) with the concentration of CaCl$_2$ and MgCl$_2$ solutions at contact angle of 40$^0$ and 110$^0$ for a nanochannel of height 8 nm and $\sigma  =  - 0.15\,\,{\rm{C/}}{{\rm{m}}^2}$.The hydrated diameters of Mg$^{2+}$ and Ca$^{2+}$ ions are considered as 0.6 nm and 0.52 nm respectively in accordance with the literature \cite{Tansel2006,Kiriukhin2002}.}
\begin{ruledtabular}
\begin{tabular}{cccc|ccc|ccc|ccc}
&\multicolumn{6}{c|}{CaCl$_2$, $a$ = 0.52 nm}&\multicolumn{6}{c}{MgCl$_2$, $a$ = 0.6 nm}\\ \hline
&\multicolumn{3}{c|}{$\theta_w=40^0$}&\multicolumn{3}{c|}{$\theta_w=110^0$}&\multicolumn{3}{c|}{$\theta_w=40^0$}&\multicolumn{3}{c}{$\theta_w=110^0$}\\ \hline
C&I$_{MD}$&I$_{Cont}$&\%&I$_{MD}$&I$_{Cont}$&\%&I$_{MD}$&I$_{Cont}$&\%&I$_{MD}$&I$_{Cont}$&\%\\
(M)&(A/m)&(A/m)&error&(A/m)&(A/m)&error&(A/m)&(A/m)&error&(A/m)&(A/m)&error\\ \hline
0.5&5.48&5.46&0.36&88.58&91.18&2.94&5.54&5.4&2.53&88.58&91.11&2.86\\ \hline
0.6&5.42&5.43&0.18&88.69&91.16&2.78&5.43&5.37&1.1&88.64&91.08&2.75\\ \hline
0.7&5.43&5.4&0.55&88.74&91.15&2.72&5.48&5.34&2.55&88.73&91.06&2.62\\ \hline
0.8&5.4&5.38&0.37&88.81&91.13&2.61&5.45&5.32&2.38&88.82&91.04&2.5\\ \hline
0.9&5.42&5.37&0.92&88.83&91.12&2.58&5.45&5.3&2.75&88.85&90.51&1.87\\ \hline
1&5.44&5.35&1.65&88.95&91.1&2.42&5.43&5.26&3.13&88.93&90.47&1.73\\
\end{tabular}
\end{ruledtabular}
\end{table*}

For low surface charge densities, low ionic concentrations, and monovalent counterions, Boltzmann distribution of ions is valid, implying no CI. However, in the other spectrum of high surface charge density or multivalent counterions, due to increased electrostatic interactions between substrate and counterions, a high counterion density exists near the substrate which can lead to CI. Lower substrate wettability also increases counterion density near the substrate, thereby leading to CI due to several reasons - (i) counterions can occupy the vacant sites left by the solvent molecules, (ii) reduced possibility of existence of hydrated ion near the substrate increases the attractive force between substrate and the counterions, and (iii) high electric field near the substrate due to the reduced solvent permittivity further increases the corresponding attractive force. In conclusion, an inverted interfacial potential, which is due to CI, can be observed even at low surface charge density provided the counterions are multivalent and substrate is hydrophobic as shown in Fig. 2(a) for the case of $\left( {{\theta _w} = {100^0}} \right)$.

\begin{figure}[hbt]
\includegraphics[scale = 0.35]{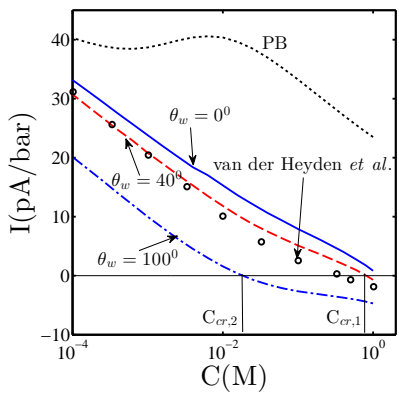}
\caption{\label{fig:fig_3}\small (color online) Variation of streaming current per unit pressure drop with the concentration of MgCl$_2	$ solution and for different contact angles in a channel of height 490 nm and $\sigma  =  - 0.15\,{\rm{C/}}{{\rm{m}}^2}$. Solid, dashed and dash-dotted lines represent results obtained for contact angles of 0$^0$, 40$^0$ and 100$^0$ respectively. Symbols represent the experimental results as reported by van der Heyden {\it et al.}\cite{vanderHeyden2006}. Dotted lines represent the results obtained via the Poisson-Boltzmann  route. The hydrated ionic diameter of Mg$^{2+}$ is taken from \cite{Tansel2006,Kiriukhin2002}.}
\end{figure}

Table~\ref{tab:table_1} depicts the streaming currents obtained from MD simulation (see \cite{SI} for the details of MD simulations and streaming current calculations) of electrokinetic transport of CaCl$_2$ and MgCl$_2$ solutions in a channel of 8 nm height, as compared against those obtained using the present model, for different contact angles.  In this context, it may be noted that CI has been experimentally observed for MgCl$_2$ and CaCl$_2$ solutions by streaming current measurements \cite{vanderHeyden2006} where a negative advective current beyond a threshold ionic concentration indicates CI, albeit in larger channels (490 nm) that may be experimentally probed unlike the ones investigated through MD simulations (8 nm). Continuum-based explanations of these experimental observations, however, have overlooked the effect of the substrate wettability on the pertinent observations \cite{Labbez2009,Gillespie2011,Storey2012}. Towards this, we obtain the streaming currents at the contact angles of 0$^0$ and 40$^0$, in accordance with the contact angles on fused silica substrates \cite{Lamb1982,Israelachvili1989}, which are shown in Fig.~\ref{fig:fig_3} along with reported experimental data (denoted by open circles) \cite{vanderHeyden2006}.

It can be seen from Table~\ref{tab:table_1} that the streaming currents obtained using the present model are in good agreement with the MD simulation results. These results upscale nicely for relatively larger channels (490 nm) as well, and match well with experimental trends \cite{vanderHeyden2006} (see Fig.~\ref{fig:fig_3}). As seen from the MD simulations and continuum predictions in Table~\ref{tab:table_1}, streaming currents increase with the contact angle for ${2h = 8\,{\rm{nm}}}$. This is due to an increase in both the solvent velocity (see \cite{SI} for MD evidence of hydrophobicity$-$induced slipping hydrodynamics) and counterion charge density near the wall with a decrease in the substrate wettability.

We note, however, that the streaming currents measured in experiments and obtained using continuum predictions reduce with the contact angle as shown in Fig.~\ref{fig:fig_3}. This is because, at experimental length scales, the slip length is much smaller than the channel height $\left( {2h = 490\,{\rm{nm}}} \right)$, thus rendering the condition at the wall to be effectively no-slip \cite{Lauga2007}. Moreover, a layer of fluid near the wall could be immobile due to the increased viscosity of the solvent which itself is due to the jamming of the counterions \cite{Bazant2009, *Bazant2009_1}. Therefore, despite an increase in the counterion charge density near the wall with an increase in the contact angle, a simultaneous increase in the width of immobile region near the wall due to crowding of counterions leads to a decrease in the streaming current with an increase in the contact angle over the reported experimental scales. In this work, the jamming layer is defined as distance over which ionic concentration is greater than a threshold concentration as depicted in the inset of Fig. 2(b). Quite naturally, the results obtained from the PB equation show gross overestimations in streaming current as compared to the reported experimental data. Even though the streaming currents obtained by using a contact angle of 0$^0$ and 40$^0$ are close to the experimental results, the most important artefact of the experimental observations $-$ negative streaming currents at high concentrations, are {\it never} observed for a contact angle of 0$^0$. On the other hand, such negative streaming currents are observed by using a contact angle of 40$^0$, thereby suggesting the focal role of substrate wettability towards consistent predictions of experimental observations. It can also be observed that the reversal of streaming currents occurs at progressively lower ionic concentrations as the contact angle increases. Similar to the influence of substrate hydrophobicity, with an increase of ionic concentration, more number of counterions get packed near the surface (as manifested through the decrease in Debye length in the mean field theory) and this can lead to CI. As the inversion of streaming current is a manifestation of CI, such an inverted streaming current can also be observed at lower ionic concentrations provided that the wettability of the substrate is less than a threshold. This is evident from Fig.~\ref{fig:fig_3}, where the threshold ionic concentration for the inversion of streaming current at a contact of 100$^0$ (C$_{cr,2}$) is lower than the threshold ionic concentration at a contact angle of 40$^0$ (C$_{cr,1}$).

To summarize, we have provided a new theoretical framework for revealing the role of substrate wettability on the electrokinetic transport through nanochannels, for a solution containing multivalent ions. We demonstrate the dual role played by substrate wettability towards affecting the electrokinetics $-$ on one hand it leads to an alteration in the ionic charge density near the wall due to solvent-substrate-ion interactions, while on the other hand it leads to a change in the current flux depending on the interplay of slip length and the characteristic length-scale of the channel. The streaming currents evaluated from such a framework are in excellent agreements with simulations and experiments, yielding a broader perspective in the inconspicuous yet decisive role of the substrate towards the electrokinetic phenomena and biochemical ion transport processes prevalent in nature and engineering.

We would like to acknowledge Mr. Chirodeep Bakli for his help in carrying out MD simulations and insightful discussions regarding the same.

\appendix
\section{\label{sec:sec1_lev1}Modifications in electrostatic potential beyond the Poisson-Boltzmann picture}
We first extend the traditional Boltzmann distribution of ions by accounting for the finite size of ions through an excluded volume contribution and the electrostatic correlations among the ions through a fluctuation potential \cite{Outhwaite1980, *Outhwaite1982, *Outhwaite1983, BariBhuiyan2004}. The key assumption behind the modeling of these terms, as given in the literature \cite{Outhwaite1980, Outhwaite1982, Outhwaite1983, BariBhuiyan2004}, is to regard the solvent as continuous uniform dielectric with a bulk dielectric constant. However, this needs to be supplemented with additional artefacts because of hydrophobic interactions. In particular, the loss of hydrogen bonds near an extended hydrophobic surface causes water to move away from these surfaces, thereby producing a wall-adjacent depletion layer \cite{Lum1999}. According to Joly {\it et al}. \cite{Joly2004}, the depletion of solvent near a hydrophobic surface gives rise to an excess effective interaction potential in addition to the electrostatic potential as ${V_{ext}} =  - {k_b}T\ln \left( {\rho \left( {\bar y} \right)/{\rho _l}} \right)$, where ${k_b}$ is the Boltzmann constant, {\it T}(= 298 K in this case) is the absolute temperature of the solvent, $\rho \left( y \right)$ is the near wall solvent density distribution, and ${\rho _l}$ is the bulk solvent density. This excess potential contribution is calculated in the present theory by expressing $\rho$ as a function of the order parameter, $\phi$, and appealing to the numerically obtained profile variation of $\phi$.

\section{\label{sec:sec2_lev1} Governing equation for EDL potential distribution}

For a binary electrolyte $\left( {{z_p}:{z_n}} \right)$, where the subscripts {\it p} and {\it n} denote the positive and negative ions respectively, the non-dimensional governing differential equations and boundary conditions are obtained using $\bar{y} = y/a,\,\,\bar{\psi}  = \beta e\psi ,\,\,{L_1} = \beta eL,\,\,{\bar{\psi _v}} = {\psi _v}\beta e/a,\,\,{\bar B_0} = {B_0}\beta e/a$, ${\bar \phi _f} = {\phi _f}\beta e$ and $\bar \chi  = \chi /{\chi _l}$ as the non-dimensional variables and introducing the dimensionless parameters as ${y_0} = {k_0}a$, $\Gamma (\bar y) = \beta {e^2}/(4\pi \varepsilon (\bar y)a)$, where $\Gamma \left( {\bar y} \right)$ is the Coulomb coupling parameter \cite{Outhwaite1980}. Due to finite ionic size, no charge exists within a distance of radius of the ion from the wall (Stern layer). Accordingly, the dimensionless Poisson equation reduces to \cite{Outhwaite1980} 
\begin{equation}
\frac{d}{{d\bar y}}\left( {\bar \varepsilon (\bar y)\frac{{d\bar \psi }}{{d\bar y}}} \right) = 0,\;\;\;{\rm{for }}\;\;\;0 \le \bar y \le 1/2
\label{eq:1},
\end{equation}
in the Stern layer, whereas in the diffuse layer, we have \cite{Outhwaite1980}
\begin{equation}
\frac{d}{{d\bar y}}\left( {\bar \varepsilon (\bar y)\frac{{d\bar \psi }}{{d\bar y}}} \right) =  - \frac{{y_0^2}}{{{z_p} - {z_n}}}[{g_{0p}} - {g_{0n}}],\;{\rm{for }}\;\bar y \ge 1/2
\label{eq:2}
\end{equation}
The corresponding wall-ion distribution functions are given as \cite{Outhwaite1980}
\begin{subequations}
\label{eq:3}
\begin{equation}
{g_{0p}} = \frac{{\rho \left( {\bar y} \right)}}{{{\rho _l}}}\zeta(\bar{y}) \left[ { - {z_p}{L_1}(\bar \psi ) - \frac{{z_p^2\Gamma (\bar y)}}{2}(F - {F_0})} \right]
\label{subeq:3a}
\end{equation}
\begin{equation}
{g_{0n}} = \frac{{\rho \left( {\bar y} \right)}}{{{\rho _l}}}\zeta(\bar{y}) \left[ {-{z_n}{L_1}(\bar \psi ) - \frac{{z_n^2\Gamma (\bar y)}}{2}(F - {F_0})} \right]
\label{subeq:3b}
\end{equation}
\end{subequations}
where  \cite{Outhwaite1980, Outhwaite1982, Outhwaite1983, BariBhuiyan2004},
\begin{subequations}
\label{eq:4}
\begin{eqnarray}
{L_1}(\bar \psi ) = \frac{F}{2}\bar \psi (\bar y + 1) + \frac{1}{2}\left[ {1 + (F - 1)(\bar y - 1/2)} \right]\bar \psi (0)\nonumber\\
+ \frac{1}{4}\left[ {(3/2 - \bar y)(\bar y - 1/2) + F({{\bar y}^2} - 5/4)} \right]\bar \psi '(0)\nonumber\\
- \frac{{(F - 1)}}{2}\int\limits_{1/2}^{\bar y + 1} {\bar \psi (\bar y')d\bar y'},\;\;{\rm{for }}\;\;1/2 \le \bar y \le 3/2,\;\;\;\;
\label{subeq:4a}
\end{eqnarray}
\begin{eqnarray}
{L_1}(\bar \psi ) = \frac{F}{2}\left[ {\bar \psi (\bar y + 1) + \bar \psi (\bar y - 1)} \right] - \frac{{(F - 1)}}{2}\int\limits_{\bar y - 1}^{\bar y + 1} {\bar \psi (\bar y')d\bar y'}\nonumber\\
{\rm{for }}\;\;\;\bar y \ge 3/2,\;\;\;\;\;\;\;\;
\label{subeq:4b}
\end{eqnarray}
\end{subequations}
\begin{subequations}
\label{eqn:5}
\begin{eqnarray}
\label{subeq:5a}
F = (4 + f{\delta _3})/[4 + \kappa a(1 + 2\bar y) + f{\delta _4}],\nonumber\\
{\rm{for }}\;\;1/2 \le \bar y \le 3/2
\end{eqnarray}
\begin{eqnarray}
\label{subeq:5b}
F = (1 + f{\delta _2})/[(1 + \kappa a)(1 - f{\delta _1})],{\rm{for }}\;\;\bar y \ge 3/2
\end{eqnarray}
\end{subequations}
\begin{subequations}
\label{eqn:6}
\begin{eqnarray}
\label{subeq:6a}
{\delta _2} = \exp [2\kappa a(1 - \bar y)]\sinh (\kappa a)/(2\kappa a\bar y)
\end{eqnarray}
\begin{equation}
\label{subeq:6b}
{\delta _1} = {\delta _2}[\kappa a\cosh (\kappa a) - \sinh (\kappa a)]/[(1 + \kappa a)\sinh (\kappa a)]
\end{equation}
\begin{eqnarray}
\label{subeq:6c}
{\delta _3} = \frac{{(1 - \exp \{ \kappa a[{{(1 + 2\bar y)}^{1/2}} - 1 - 2\bar y]\} )}}{{\kappa a\bar y}}\nonumber\\
+ \frac{{\left( {1 - 2\bar y + {{(1 + 2\bar y)}^{1/2}}} \right)}}{{\bar y}}
\end{eqnarray}\hfill
\begin{equation}
\label{subeq:6d}
{\delta _4} = {\delta _3} - (1/\bar y)(1 + \exp \{ \kappa a[{(1 + 2\bar y)^{1/2}} - 1 - 2\bar y]\} )
\end{equation}
\end{subequations}
\begin{widetext}
\begin{eqnarray}
\label{eqn:7}
\zeta({\bar y}) = H\left( {\bar y - {\textstyle{1 \over 2}}} \right) \exp\left\{ { - 2\pi {a^3}\int\limits_{h/a}^{\bar y}{d{\bar y}_1}\left(
\begin{array}{c}
n_p^0\!\!\!\!\!\!\int\limits_{\max (1/2,{{\bar y}_1} - 1)}^{{{\bar y}_1} + 1}\!\!\!\!\!\!\!\!\!\!\!\!\!\!\!\! {(\bar Y - {{\bar y}_1}){g_{0p}}(\bar Y)\exp [ - {z_p}{{\bar \phi }_f}({{\bar y}_1},\bar Y)]d\bar Y } \\
+n_n^0\!\!\!\!\!\!\!\!\!\!\!\int\limits_{\max (1/2,{{\bar y}_1} - 1)}^{{{\bar y}_1} + 1} \!\!\!\!\!\!\!\!\!\!\!\!\!\!\!\!{(\bar Y - {{\bar y}_1}){g_{0n}}(\bar Y)\exp [-{z_n}{{\bar \phi }_f}({{\bar y}_1},\bar Y)]d\bar Y}
\end{array}\right)} \right\}\;
\end{eqnarray}
\begin{eqnarray}
\label{eqn:8}
{\bar \phi _f}({\bar y_1},\bar Y) = {\bar B_0}({\bar y_1})\left\{
{\exp ( - \kappa a) + \left[ {\frac{f}{{{{(1 + 4{{\bar y}_1}\bar Y)}^{1/2}}}}} \right]\exp [ - \kappa a{{(1 + 4{{\bar y}_1}\bar Y)}^{1/2}}]}
\right\},
\end{eqnarray}
where $f = \left( {\varepsilon  - {\varepsilon _W}} \right)/\left( {\varepsilon  + {\varepsilon _W}} \right)$, ${\varepsilon _W}$ being the permittivity of the wall. In this study, we assume, ${\varepsilon _W} = \varepsilon$, so that $f = 0$ \cite{Outhwaite1980, *Outhwaite1982, *Outhwaite1983, BariBhuiyan2004}.
\begin{subequations}
\label{eqn:9}
\begin{eqnarray}
\label{subeq:9a}
{\bar B_0}\left( {\bar y} \right) = \left( {{{\bar \psi }_v}/\pi } \right){\left\{ \begin{array}{l}
2\left[ {\left( {1 + \kappa a\bar y + \kappa a} \right)\exp \left( { - \kappa a} \right) + \exp \left( { - \kappa a\bar y} \right)} \right] - \\
\left( {f/\kappa a\bar y} \right)\left[ {2\kappa a\bar y\exp \left( { - \kappa a\bar y} \right) + \left( {1 + \kappa a} \right)\exp \left( { - \kappa a} \right)\left( {\exp \left( { - 2\kappa a\bar y} \right) - 1} \right)} \right]
\end{array} \right\}^{ - 1}},{\rm{for }}\;\;1/2 \le \bar y \le 1
\end{eqnarray}
\begin{equation}
\label{subeq:9b}
{\bar B_0}(\bar y) = \left\{ {\exp (\kappa a)/[4\pi (1 + \kappa a)(1 - f{\delta _{11}})]} \right\}{\bar \psi _v},\;\;{\rm{for }}\;\;\;\bar y \ge 1
\end{equation}
\end{subequations}
\begin{subequations}
\begin{eqnarray}
{\bar \psi _v} = \pi \left[ \begin{array}{l}
(2\bar y - 1)\bar \psi (1/2) + (\bar y + 1/2)(\bar y - 3/2)\bar \psi '(1/2)\\
 + 2\bar \psi (\bar y + 1) - 2\int\limits_{1/2}^{\bar y + a} {\bar \psi (\bar y')d\bar y'} 
\end{array} \right],\;\;\;{\rm{for }}\;\;\;1/2 \le \bar y \le 3/2
\end{eqnarray}
\begin{equation}
{\bar \psi _v} = 2\pi \left[ {\bar \psi (\bar y + 1) + \bar \psi (\bar y - 1) - \int\limits_{\bar y - 1}^{\bar y + 1} {\bar \psi (\bar y')d\bar y'} } \right],\;\;\;{\rm{for}}\;\;\;\bar y \ge 3/2
\end{equation}
\end{subequations}	
\begin{equation}
{\delta _{11}} = \frac{{\left( {\kappa a\cosh \left( {\kappa a} \right) - \sinh \left( {\kappa a} \right)} \right)\exp \left[ {\kappa a\left( {1 - 2\bar y} \right)} \right]\sinh \left( {\kappa a} \right)}}{{2\kappa \bar ya\left( {1 + \kappa a} \right)\sinh \left( {\kappa a} \right)}}
\end{equation}
\end{widetext}
The boundary conditions used to solve non-dimensional governing differential equations for  $\bar{\psi}$ with $\bar y \in \left[ {0,h/a} \right]$ are : (i) continuity of $\bar{\psi}$ and $\bar \varepsilon {\textstyle{{d\bar \psi } \over {d\bar y}}}$ at $\bar y = 1/2$, (ii) known surface charge density at the wall, $\sigma ea/\left( {{\varepsilon _l}{k_B}T} \right) =  - {\left. {\left( {\bar \varepsilon \left( {\bar y} \right)d\bar \psi /d\bar y} \right)} \right|_{\bar y = 0}}$ and (iii) symmetry boundary condition at the centerline of the channel, ${\left. {\left( {d\bar \psi /d\bar y} \right)} \right|_{\bar y = h/a}} = 0$, where $h$ is the half channel height. We reiterate here that the $\bar{\varepsilon}$ appearing above is not constant; rather, it is determined by locally interpolating between the vapor and liquid permittivity as determined by the spatially varying order parameter. Mathematically, $\bar{\varepsilon} = \frac{\varepsilon_v}{\varepsilon_l}\frac{1+\phi}{2}+\frac{1-\phi}{2}$. These coupled, non-dimensional governing differential equations are numerically solved using commercial software COMSOL Multiphysics 4.4.

\section{\label{sec:sec3_lev1}MD Simulation Details}
The model system for MD simulation consists of a parallel plate geometry with a channel height of 8 nm and planar wall dimensions of 5 nm$\times$5 nm. Periodic boundary conditions are applied in  X, Y and Z; X being the axial direction. The channel is filled with 6936 water molecules (SPC/E model) and the number of ions is determined by the bulk concentration and overall electroneutrality of the system. After energy minimization, the system was equilibrated for 12 ns, followed by non-equilibrium simulation for another 12 ns. All the simulations are performed with time step of 1 fs. The wall molecules are kept fixed at their respective lattice positions. The fluid (water and ions) is actuated by applying a uniform acceleration in the X direction. Wall wettability is altered by tuning the Lennard-Jones (LJ) parameters of the hetero-nuclear potential between the wall and the oxygen of water molecule. The temperature is kept constant at 300 K by a Noose-Hoover thermostat  \cite{Berendsen1995} and the long range electrostatic interactions are calculated by Particle Mesh Ewald \cite{Berendsen1995}. The ion distribution and velocity profiles are obtained by suitable binning and the streaming current is calculated using the methodology discussed in section VI.

\section{\label{sec:sec4_lev1}Evidence of Slip from MD based velocity data}
\begin{figure}[hbt]
\includegraphics[scale=0.4]{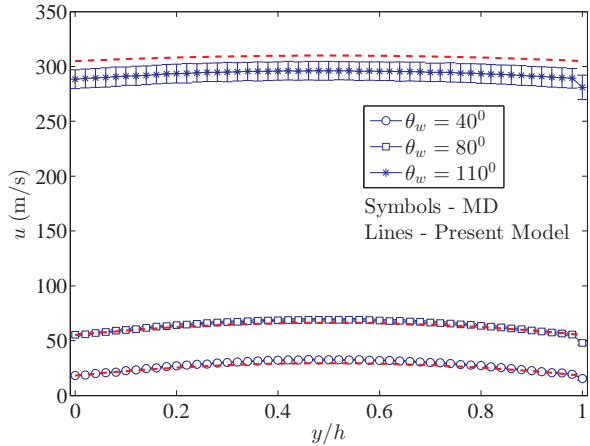}
\caption{\label{fig:fig_S1}\small (color online) Mean and standard deviation of the molecular velocity profiles (symbols) in a channel of height 8 nm and surface charge density of -0.15 C/m$^2$, obtained from the MD simulation data, are shown at contact angles of 40$^0$, 80$^0$, and 110$^0$. The dashed lines represent the fitted Poiseuille flow profiles at the corresponding contact angles. Note that the deviation from the mean velocity at the contact angles of 40$^0$ and 80$^0$ is imperceptibly small of the fitted Poiseuille profile.}
\end{figure}
Fig.~\ref{fig:fig_S1} depicts the molecular velocity data obtained in a channel of height 8 nm, with surface charge density of -0.15 C/m$^2$, at contact angles of 40$^0$, 80$^0$ and 110$^0$ respectively. This MD data is matched with a Poiseuille flow velocity profile for which the pressure gradient and slip length is obtained from the MD data. The solvent viscosity is obtained by matching the velocity gradient at the wall for both MD and continuum predictions, as was also done in the literature \cite{Joly2004, Joly2006}. Navier slip boundary condition at the channel walls, ${\left. {u} \right|_{y = 0}} = {l_s}{\left( {\frac{{\partial u}}{{\partial y}}} \right)_{y = 0}}$ is utilized for obtaining the corresponding Poiseuille flow velocity profile, where ${l_s}$ is the slip length. As seen from Fig.~\ref{fig:fig_S1}, both the slip velocity and slip length increase with the contact angle. Since for a given wettability, slip phenomenon appears to be more pronounced for channels with reduced dimensions, one can safely remark that the maximum slip length for the investigated experimental scenario (2$h$ $\approx$ 490 nm) must be lower than the same for the investigated MD scenario (2$h$ $\approx$ 8 nm). As the maximum slip length at the highest contact angle investigated is much smaller than the characteristic channel dimension over the probed experimental scales ($\approx$ 490 nm), the no slip boundary condition remains justified over the reported experimental scales, despite the emergence of slipping hydrodynamics for the corresponding physical scales probed through MD simulations ($\approx$ 8 nm)
\section{\label{sec:sec5_lev1}Near-wall density fluctuations from MD data}
\begin{figure}[hbt]
\includegraphics[scale=0.4]{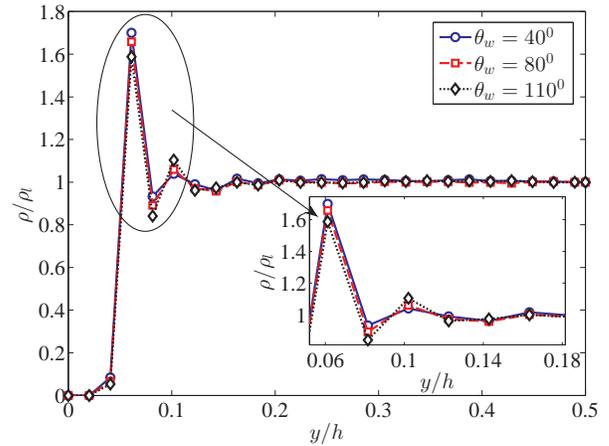}
\caption{\label{fig:fig_S2}\small (color online) Normalized number density of water molecules across the half channel height of a channel of height 8 nm, surface charge density of -0.15 C/m$^2$, at different contact angles. The circles, diamonds and square symbols represent the data at contact angles of 40$^0$, 80$^0$ and 110$^0$ respectively.}
\end{figure}
In order to elucidate the influence of substrate wettability on the slip velocity or the velocity profile, we plot the fluid density profile across the half channel height. Fig.~\ref{fig:fig_S2} shows the plot of the normalized density of water molecules against the channel height at contact angles of 40$^0$, 80$^0$ and 110$^0$. For low values of contact angles $\left( {{\theta _w} = {{40}^0}} \right)$, the number density of water molecules exhibits a hump near the surface, which progressively gets attenuated due to the bulk thermal motion as seen from Fig.~\ref{fig:fig_S2}. At contact angles of 110$^0$, the surface becomes more hydrophobic, so that the water density near the surface becomes smaller than that at lower contact angles as shown in Fig.~\ref{fig:fig_S2}. This decreasing near-wall water density hallmarks the increasing interfacial slip as depicted in Fig.~\ref{fig:fig_S1}.

\section{\label{sec:sec6_lev1}Calculation of Streaming Currents}
Substrate wettability not only affects the electrostatics but also significantly influences the associated hydrodynamics, explicitly through the density variations and implicitly through its effects on electrostatics, as experimentally observed through streaming current. Streaming current is defined as the advective pressure driven current of the mobile ions under the constraint of a net zero current (sum of advection current and electromigration current) in the system \cite{Hunter1981}. The streaming current is thus given by $I = Wa\int\limits_0^{2h/a} {{\rho _e}\left( {\bar y} \right)u\left( {\bar y} \right)d\bar y} $, where {\it W} is the channel width, ${\rho _e}\left( {\bar y} \right)$ is the local charge density and $u\left( {\bar y} \right)$ is the fluid velocity profile. Over length scales in the purview of MD simulations, the velocity profiles are calculated as explained in the Section IV. At experimental length scale of $2h=490$ nm, the slip length being much smaller than the channel height $\left( {2h = 490\,{\rm{nm}}} \right)$, the no slip boundary condition at the wall may still be considered to be valid \cite{Lauga2007}. Moreover, a layer of fluid near the wall is likely to be immobile due to the increased viscosity of the solvent which itself is due to the jamming of the counterions, a phenomenon termed as Charge Induced Thickening \cite{Bazant2009, Bazant2009_1}. Therefore, over such experimental length scales, the velocity profile is given by,
\begin{equation}
u\left( {\bar y} \right)=\left\{
\begin{array}{c}
0,\;\;\;\; \,{\rm{for }}\;\;0 < \bar y < \gamma\\
\frac{{{a^2}}}{\mu }\frac{{dP}}{{dx}}\left( {\frac{{{{\bar y}^2} - {\gamma ^2}}}{2} - \left( {\bar y - \gamma } \right)\frac{h}{a}} \right) + \frac{{{E_s}{k_b}T{\varepsilon _l}}}{{e\mu }}\int\limits_\gamma ^{\bar y} {\bar \varepsilon \frac{{d\bar \psi }}{{d\eta }}d\eta },\\
\;\;\;\;\;\;\;\;\;\;\;\;\;{\rm{for }}\;\;\gamma  < \bar y < h/a
\end{array}\right\},
\end{equation}
where $\gamma$ is the non-dimensional distance over which the concentration exceeds the threshold concentration as defined in the main text (see the inset of Fig. 2(b) in the main text for the details about calculation of threshold concentration). Here the first term represents the flow due to applied pressure gradient, $-dP/dx$. The second term represents the flow due to induced streaming field $\left( {{E_s}} \right)$ which satisfies an overall current electro neutrality \cite{Hunter1981}
\begin{widetext}
\begin{eqnarray}
{I_{{\rm{net}}}} = \underbrace {ae{n^0}\int_\gamma ^{2h/a} {\left( {{g_{0p}} - {g_{0n}}} \right)u\left( {\bar y} \right)d\bar y} }_{{\rm{Streaming\;\;Current}}}+ \underbrace {\left( {{E_s}a\lambda /2} \right)\int_0^{2h/a} {\left( {{g_{0p}} + {g_{0n}}} \right)} d\bar y}_{{\rm{Conduction\;\;Current}}} = 0,
\end{eqnarray} leading to
\begin{equation}
{E_s} =  - \frac{{\frac{{{a^2}e{n^0}}}{\mu }\frac{{dP}}{{dx}}\int\limits_\gamma ^{h/a} {({g_{0p}} - {g_{0n}})\left( {\frac{{{{\bar y}^2} - {\gamma ^2}}}{2} - \left( {\bar y - \gamma } \right)\frac{h}{a}} \right)d\bar y} }}{{\frac{\lambda }{2}\int\limits_0^{h/a} {({g_{0p}} + {g_{0n}})d\bar y}  + \frac{{{\varepsilon _l}{n^0}{k_b}T}}{\mu }\int\limits_\gamma ^{h/a} {({g_{0p}} - {g_{0n}})\left\{ {\int\limits_\gamma ^{\bar y} {\bar \varepsilon \frac{{d\bar \psi }}{{d\eta }}d\eta } } \right\}d\bar y} }}
\end{equation}
\end{widetext}
where $\mu$ is the dynamic viscosity of the liquid water and $\lambda$ is the electrical conductivity of MgCl$_2$ solution obtained from \cite{Phang1980}. The sreaming currents per unit width from the MD simulations is calculated using ${I_1} = \int_0^{2h} {{\rho _{e1}}\left( y \right){u_1}\left( y \right)dy}$, where the variables ${\rho _{e1}}\left( y \right)$ and ${u_1}\left( y \right)$ represent local charge density and velocity calculated from the MD simulations through post processing.
\nocite{*}

\bibliography{draft_references}

\end{document}